\documentclass[a4paper,11pt]{article}

\newcommand{\be}{\begin{equation}}
\newcommand{\ee}{\end{equation}}
\newcommand{\bea}{\begin{eqnarray}}
\newcommand{\eea}{\end{eqnarray}}
\newcommand{\ena}{\end{eqnarray}}

\def\half{\frac{1}{2}}

\usepackage{graphicx,amssymb,amsmath,bm,latexsym,color}
\makeatletter
\@addtoreset{equation}{section}

\makeatother
\newcommand{\vs}[1]{\vspace{#1 mm}}
\newcommand{\hs}[1]{\hspace{#1 mm}}
\renewcommand{\a}{\alpha}
\renewcommand{\b}{\beta}
\renewcommand{\c}{\gamma}
\renewcommand{\d}{\delta}
\newcommand{\e}{\epsilon}

\renewcommand{\t}{\theta}

\newcommand{\pa}{\partial}
\newcommand{\nn}{\nonumber\\}
\newcommand{\p}[1]{(\ref{#1})}

\newcommand{\td}{\tilde d}
\newcommand{\sr}{\hspace{-.67em}/}
\newcommand{\vect}[1]{\!\!\!\mbox{ \boldmath $#1$}}

\begin{document}

\begin{titlepage}

\begin{flushright}
KU-TP 029 \\
\end{flushright}

\vskip .5in

\begin{center}

{\Large\bf Supersymmetric Intersecting Branes \\
in Time-dependent Backgrounds}
\vskip .5in

{\large
Kei-ichi Maeda,$^{a,b,}$\footnote{E-mail address: maeda``at"waseda.jp}
Nobuyoshi Ohta,$^{c,}$\footnote{E-mail address: ohtan``at"phys.kindai.ac.jp}
Makoto Tanabe$^{a,}$\footnote{E-mail address:
tanabe``at"gravity.phys.waseda.ac.jp}
and
Ryo Wakebe$^{a,}$\footnote{E-mail address:
wakebe``at"gravity.phys.waseda.ac.jp}
}\\
\vs{10}
$^a${\em Department of Physics, Waseda University,
Shinjuku, Tokyo 169-8555, Japan} \\
$^b${\em Waseda Research Institute for Science and Engineering,  Shinjuku, Tokyo 169-8555, Japan}\\
$^c${\em Department of Physics, Kinki University,
Higashi-Osaka, Osaka 577-8502, Japan} \\

\vskip .2in \vspace{.3in}

\begin{abstract}

We construct a fairy general family of supersymmetric solutions in
time- and space-dependent backgrounds in general supergravity theories.
One class of the solutions are intersecting brane solutions with factorized
form of time- and space-dependent metrics, the second class are brane
solutions in pp-wave backgrounds carrying spacetime-dependence,
and the final class are the intersecting branes with more nontrivial
spacetime-dependence, and their intersection rules are given.
Physical properties of these solutions are discussed, and the relation
to existing literature is also briefly mentioned.
The number of remaining supersymmetries are identified for various
configurations including single branes, D1-D5, D2-D6-branes with
nontrivial dilaton, and their possible dual theories are briefly discussed.

\end{abstract}

\end{center}

\vfill

\end{titlepage}
\tableofcontents

\newpage
\setcounter{page}{1}
\section{Introduction}

The understanding of the fundamental nature and quantum properties
of spacetime is one of the most important questions in theoretical
physics. An example of such problems is the spacetime singularities
that general relativity predicts. A well-known one is the big-bang
singularity of a time-dependent spacetime, where general relativity breaks
down. One needs a quantum theory of gravity to understand physics close to
the singularity. String theory is one of the most promising candidate for
such a theory. Although we know some static solutions in string theory, e.g.
products of Minkowski space and compact manifolds, these static spacetimes
are not so much useful in clarifying the dynamics of string theory in the
strong curvature regime or near the singularities. Therefore it is necessary
to understand string theory on time-dependent backgrounds. Unfortunately
time-dependent backgrounds are difficult to work with in string theories in general,
though some special cases are analyzed~\cite{time1}-\cite{null}.

Recently a model of big-bang cosmology has been proposed in matrix string theory
based on AdS/CFT correspondence which is a powerful nonperturbative formulation
of string theory~\cite{bb1}. This corresponds to a simple time-dependent solution
of supergravities which are the low-energy effective theories of string theories
that preserves 1/2 supersymmetry in ten dimensions, with the light-like linear
dilaton background, and various extensions have been considered~\cite{bb2}-\cite{BPRW}.
As is usual in AdS/CFT correspondence, supersymmetry is expected
to play an important role. The existence of supersymmetry allows us to better
control the behaviors of the solutions in string/supergravity backgrounds and
the quantum and nonperturbative properties of the field theories. Therefore there has
been much interest in time-dependent supersymmetric solutions of
string/supergravity theories.
For a detailed review of the big-bang models in string theory, see~\cite{BC}.

On the other hand, D-branes can probe the nonperturbative dynamics of the string
theory and they have been used to study various duality aspects of string theory.
It is thus interesting to find if we can have such brane solutions
in time-dependent backgrounds with time-dependent dilaton.
In fact, D3-brane solutions have been found and discussed in
\cite{CH,DMNT1} and other single brane solutions in \cite{NP,NPS}.
A systematic derivation of the general brane solutions in the pp-wave
backgrounds has been given in \cite{OPS}.
It is also interesting to study intersecting brane systems because
it is known that some such configurations can describe the standard model
of particle physics. More recently, gauge theories on D-branes are examined
to gain into the dynamical supersymmetry breaking~\cite{DSB}.
These solutions are interesting since the metrics of these solutions depend
on both space and time, but the dependence is restricted to the product form
of these functions. The question then naturally arises if there are solutions
with more general dependence on space and time and if such solutions can
give more physical insight.

In this paper, we investigate more general time-dependent supersymmetric
solutions in supergravity theories in ten and eleven dimensions in order to
understand the nature of spacetime.
In sect.~2, we derive brane solutions in general supergravities with dilaton
and forms of arbitrary ranks in spacetime-dependent backgrounds.
In sect.~3, we give time-dependent solutions restricted to those with
time-independent harmonic functions. All known solutions belong to this class of
solutions, but our solutions are more general. We clarify the relation of our
solutions and the known ones.
In sect.~4, we give more general solutions with time-dependent harmonic functions
for one brane and two intersecting branes.
These are new solutions and the physical properties of these solutions
including spacetime and asymptotic structures are discussed in sect.~5.
In sect.~6, we show that these solutions have unbroken supersymmetry,
and identify the amount of remaining supersymmetries.
Sect.~7 is devoted to conclusions and discussions.

\section{Time-dependent brane system in supergravity}

The low-energy effective action for the supergravity system
coupled to dilaton and $n_A$- form field strength is given by
\bea
I = \frac{1}{16 \pi G_D} \int d^D x \sqrt{\mathstrut-g} \left[
 R - \frac12 (\pa \Phi)^2 - \sum_{A=1}^m \frac{1}{2 n_A!} e^{a_A \Phi}
 F_{n_A}^2 \right],
\label{action}
\ena
where $G_D$ is the Newton constant in $D$
dimensions and $g$ is the determinant of the metric. The last term
includes both RR and NS-NS field strengths, and $a_A = \frac12
(5-n_A)$ for RR field strength and $a_A = -1$ for NS-NS 3-form.
In the eleven-dimensional supergravity, there is a four-form and no dilaton.
We put fermions and other background fields to be zero.

{}From the action (\ref{action}), one can derive the field
equations
\bea
R_{\mu\nu} = \frac12 \pa_\mu
\Phi \pa_\nu \Phi + \sum_{A} \frac{1}{2 n_A!}
 e^{a_A \Phi} \Biggl[ n_A \left( F_{n_A}^2 \right)_{\mu\nu}
 - \frac{n_A -1}{D-2} F_{n_A}^2 g_{\mu\nu} \Biggr],
\label{Einstein}
\ena
\bea
\Box \Phi = \sum_{A} \frac{a_A}{2 n_A!}
e^{a_A \Phi} F_{n_A}^2,
\label{dila}
\ena
\bea
\pa_{\mu_1} \left(
\sqrt{\mathstrut- g} e^{a_A \Phi} F^{\mu_1 \cdots \mu_{n_A}} \right) = 0
\,,
\label{field}
\ena
 where $F_{n_A}^2$ denotes $F_{\mu_1 \cdots \mu_{n_A}}F^{\mu_1 \cdots \mu_{n_A}}$
and $(F_{n_A}^2)_{\mu\nu}$
 denotes $F_{\mu\rho\cdots\sigma}F_\nu^{~\rho\cdots\sigma}$.

 The Bianchi identity for the form field is given by
\bea
\pa _{[\mu} F_{\mu_1 \cdots \mu_{n_A}]} =0.
\label{bianchi}
\ena

In this paper we  assume  the following metric form:
\bea
ds_D^2 = e^{2 \Xi(u,r)} \left[-2dudv + K(u,y^\a, r) du^2\right] +
\sum_{\a=1}^{d-2} e^{2 Z_\a(u,r)} (dy^\a)^2 +e^{2B(u,r)}\left(dr^2 + r^2
d\Omega_{\td +1}^2\right),
\label{met}
\ena
where $D=d+\tilde d+2$, the coordinates $u$, $v$ and $y^\a, (\a=1,\ldots, d-2)$
parameterize the $d$-dimensional worldvolume  where
the branes belong, and the
remaining $\tilde d + 2 $ coordinates $r$ and angles
are transverse to the brane worldvolume,
$d\Omega_{\tilde d+1}^2$ is the line element of the $(\td+1)$-dimensional
sphere. Note that $u$ and $v$ are null coordinates.
The metric components $\Xi, Z_\a, B$ and the dilaton $\Phi$ are assumed
to be functions of $u$ and $r$, whereas $K$ depends on $u,y^\a$ and $r$.
Our ansatz includes more general solutions than those in \cite{CH,DMNT1},
which consider only single D3-brane solutions with the metrics of product form
of time- and space-dependent factors; ours allows intersecting branes as well as
more general spacetime dependence.

For the field strength backgrounds, we take
\bea
F_{n_A} = E_A'(u,r) \, du \wedge dv \wedge dy^{\a_1} \wedge \cdots \wedge
dy^{\a_{q_A -1}} \wedge dr,
\label{eleb}
\ena
where $n_A = q_A +2$.
Throughout this paper, the dot and prime denote derivatives with
respect to $u$ and $r$, respectively. The ansatz~\p{eleb} means that
we have an electric
background.
We could, however, also include magnetic background
in the same form as the electric one with the
replacement
\bea
g_{\mu\nu} \to g_{\mu\nu}, \quad F_{n} \to e^{a
\Phi}*\! F_{n}, \quad \Phi \to - \Phi. \label{sdual}
\ena
This is due to the S-duality symmetry of the original system~\p{action}.
So we do not have to consider it separately.

With our ansatz, the Einstein equations~\p{Einstein} reduce to
\bea
&& \Xi'' + \Big( U' + \frac{\tilde d+1}{r} \Big)\Xi'
= \sum_A \frac{D-q_A-3}{2(D-2)} S_A (E'_A)^2,
\label{fe1} \\
&& \sum_{\a=1}^{d-2} \ddot Z_\a + (\tilde d + 2) \ddot B +
\sum_{\a=1}^{d-2} {\dot Z_\a}^2 + (\tilde d + 2) {\dot B}^2
- 2 \dot \Xi \left[ \sum_{\a=1}^{d-2} \dot Z_\a + (\tilde d + 2) \dot B
\right] \nn
&& +\; \frac12 \sum_{\a=1}^{d-2} e^{2(\Xi-Z_\a)}\pa^2_\a K
 + e^{2(\Xi- B)} \Bigg[ K \Xi'' + \frac12 K''
+ \Big(\Xi' K + \frac12 K'\Big) \Big(U'+\frac{\tilde d + 1}{r}\Big) \Bigg] \nn
&& \hs{20}= \sum_A\frac{D-q_A-3}{2(D-2)} e^{2(\Xi-B)} K S_A (E'_A)^2
- \frac12 (\dot \Phi)^2,
\label{fe2}
\\
&& \dot \Xi' + \sum_{\a=1}^{d-2} \dot Z_\a'+(\tilde d +1)\dot B'
- \left[ \sum_{\a=1}^{d-2}\dot Z_\a +(\td +2)\dot B\right] \Xi'
- \dot B \sum_{\a=1}^{d-2} Z_\a' + \sum_{\a=1}^{d-2} \dot Z_\a Z_\a'
= -\frac12 \dot \Phi \Phi',~~~~~~~~
\label{fe3} \\
&& Z_\a'' + \Big( U' + \frac{\tilde d+1}{r}\Big) Z_\a'
= \sum_A \frac{\d^{(\a)}_A}{2(D-2)} S_A (E'_A)^2,
\label{fe4}
\\
&& U''+B'' - \Big( 2\Xi' +\sum_{\a=1}^{d-2} Z_\a'
- \frac{\tilde d+1}{r}\Big) B' +2(\Xi')^2 + \sum_{\a=1}^{d-2} (Z_\a')^2 \nn
&& \hs{20}= - \frac12 (\Phi')^2 + \sum_A\frac{D-q_A-3}{2(D-2)}S_A (E'_A)^2,~~~~~
\label{fe5}
\\
&& B'' + \Big(U'+\frac{\tilde d+1}{r}\Big) B' +\frac{U'}{r}
= - \sum_A\frac{q_A+1}{2(D-2)}S_A (E'_A)^2,
\label{fe6}
\ena
where $U$, $S_A$ and $\d_A^{(\a)}$ are defined by
\bea
U &\equiv& 2\Xi+ \sum_{\a=1}^{d-2} Z_\a +\tilde d B\,,
\\
S_A &\equiv& \exp\left[\e_A a_A\Phi - 2\left(2\Xi
+\sum_{\a \in q_A} Z_\a \right)
\right],
\label{sa}
\ena
and
\bea
\d_A^{(\a)} = \left\{
\begin{array}{l}
D-q_A-3 \\
-(q_A+1)
\end{array}
\right. \hs{5} {\rm for} \hs{3} \left\{
\begin{array}{l}
y^\a \mbox{   belonging  to $q_A$-brane} \\
{\rm otherwise}
\end{array}
\right. ,
\ena
respectively, and $\e_A= +1 (-1)$ is for electric (magnetic) backgrounds.
The sum of $\a$ in Eq. (\ref{sa})
runs over the $q_A$-brane components
in the $(d-2)$-dimensional $y^\a$-space, for example
\bea
\sum_{\a \in q_A} Z_\a=\sum_{\a_A=1}^{q_A-1}Z_{\a_A}
\,.
\ena
Eqs. (\ref{fe1}), $\cdots$, (\ref{fe5}) and (\ref{fe6})
 are the $uv, uu, ur, \a\b, rr$ and $ab$ components of the
Einstein equations (\ref{Einstein}), respectively.
The dilaton equation~\p{dila} and  the
equations for the form field~\p{field} and \p{bianchi}
yield
\bea
e^{-U} r^{-(\tilde d+1)} (e^{U} r^{\tilde d+1} \Phi')'
&=& -\frac12 \sum_A\e_A a_A S_A (E'_A)^2,
\label{dil} \\
\Big( r^{\tilde d+1}e^{U} S_A E_A' \Big)' &=& 0,
\label{fe7} \\
\Big( r^{\tilde d+1}e^{U} S_A E_A' \Big)^{\centerdot} &=& 0.
\label{fe8}
\ena

We assume that $U$ is independent of $r$  but depends only on $u$.
In the case of static spacetime, it is known that under this condition
($U$ is constant in case of no dependence on $u$),
all the supersymmetric intersecting brane solutions   have been derived~\cite{NO}.
If this condition is relaxed, one may get more general non-BPS
solutions~\cite{MO}, but here we are interested in the BPS solutions.
We extend them to the time-dependent case.

{}From Eqs.~\p{fe7} and \p{fe8}, we learn that
\bea
r^{\td+1} e^U S_A E_A' = c_A,
\label{norm}
\ena
is a constant. Combined with Eq.~\p{dil}, we then get
\bea
\Phi' = -\frac12 \sum_A \e_A a_A \frac{c_A \tilde E_A}{r^{\td+1}},
\label{p}
\ena
where we have defined
\bea
\tilde E_A = e^{-U} E_A.
\ena
Similarly from Eqs.~\p{fe1}, \p{fe4}, \p{fe6}, we find
\bea
\Xi' &=& \sum_A \frac{D-q_A-3}{2(D-2)} \frac{c_A \tilde E_A}{r^{\td+1}}, \nn
Z_\a' &=& \sum_A \frac{\d_A^{(\a)}}{2(D-2)}
\frac{c_A \tilde E_A}{r^{\td+1}},
\nn
B' &=& -\sum_A \frac{q_A+1}{2(D-2)} \frac{c_A \tilde E_A}{r^{\td+1}}.
\label{first}
\ena
Note that there is no integral constant in the right hand sides of \p{p}
and \p{first}. This is related to the BPS condition.

Substituting these into \p{fe5}, we get
\bea
\sum_{A,B}\Big[ \frac{c_A}{2} M_{AB} + r^{\td +1}
\left(\frac{1}{\tilde E_A}\right)'
\d_{AB}\Big] \frac{c_B}{2}\frac{\tilde E_A \tilde E_B}{r^{2\td+2}}=0,
\label{norm1}
\ena
where
\bea
M_{AB} = \frac{2(D-q_A-3)(D-q_B-3)}{(D-2)^2}+\sum_{\a=1}^{d-2} \frac{\d_A^{(\a)} \d_B^{(\a)}}{(D-2)^2}
+ \tilde d \frac{(q_A+1)(q_B+1)}{(D-2)^2} + \frac12 \e_A a_A \e_B a_B.
\ena
We require that all the branes be independent, and so $E_A$ are independent
functions. We thus learn from Eq.~\p{norm1} that
\bea
\frac{c_A}{2} M_{AB} + r^{\td +1}\left(\frac{1}{\tilde E_A}\right)'\d_{AB}=0,
\label{norm2}
\ena
the off-diagonal part of which is $M_{AB}=0$ for $A \neq B$. As shown in
Ref.~\cite{NO,OP}, this condition leads to the intersection rules for two branes.
If $q_A$-brane and $q_B$-brane intersect over ${\bar q}\; (\leq q_A, q_B)$
dimensions, this gives
\bea
{\bar q} = \frac{(q_A+1)(q_B+1)}{D-2}-1 - \frac12 \e_A a_A \e_B a_B.
\label{int}
\ena
The rule (\ref{int}) tells us that
D1-branes with $a_1=1$ can intersect with D3-brane with $a_3=0$ on
a point $(\bar q=0)$ and with D5-brane with $\epsilon_5 a_5=-1$ over
a string $(\bar q=1)$, and D5-brane can intersect with
D5-brane over 3-brane $(\bar q=3)$, in agreement with Refs.~\cite{AEH,PT,T}.

The second term in \p{norm2} must be constant. This, in particular, means
\bea
H_A = \sqrt{\frac{2(D-2)}{\Delta_A}} \frac{1}{\tilde E_A}\,,
\label{fs}
\ena
is a harmonic function
\bea
(r^{\tilde d+1} H_A')'=0, \quad
(r^{\tilde d+1} H_A')^{{}^\centerdot}=0,
\label{harm}
\ena
where we have defined
\bea
\Delta_A = (D-q_A-3)(q_A+1) +\frac{D-2}{2}a_A^2.
\ena
Note, however, that the condition~\p{harm} allows $u$-dependent term
\bea
H_A = h_A(u) + \frac{Q_A}{r^{\td}},
\label{harm1}
\ena
where $h_A$ is an arbitrary function of $u$ and $Q_A$ is a constant.
This class of solutions generalize those discussed in \cite{OP}.
They are also similar to those discussed in \cite{KU}
though time-dependence is taken differently.

Using \p{fs} in \p{first}, we find
\bea
\Xi &=& - \sum_A \frac{D-q_A-3}{\Delta_A} \ln H_A + \xi(u), \nn
Z_\a &=& - \sum_A \frac{\d_A^{(\a)}}{\Delta_A} \ln H_A + \zeta_{\a}(u),
\nn
B &=& \sum_A \frac{q_A+1}{\Delta_A} \ln H_A + \beta(u), \nn
\Phi &=& \sum_A \e_A a_A \frac{D-2}{\Delta_A} \ln H_A + \phi(u),
\label{sol}
\ena
where $\xi, \zeta_{\a}, \beta, \phi$ are functions of $u$ only.
It follows from the definition and the solutions~\p{sol} that $U$ reduces to
\bea
U = 2 \xi(u) + \sum_{\a=1}^{d-2} \zeta_{\a}(u) + \tilde d \beta(u),
\ena
consistent with our ansatz that $U$ depends only on $u$.

The condition~\p{fe7} and \p{fe8} or \p{norm}, combined
with the definition~\p{sa}
and the solution, gives
\bea
&&\e_A a_A \phi
+ 2 \sum_{\a \in\kern-0.4em / q_A} \zeta_{\a} + 2 \td \beta=0,
\label{cond1}
\\
&&
c_A = \td Q_A \sqrt{\frac{2(D-2)}{\Delta_A}}.
\ena
We then find that
$
M_{AB}=\frac{\Delta_A}{D-2} \d_{AB}.
$ 
It turns out that using the intersection rules, the condition~\p{fe3} is
reduced to
\bea
(\dot H_A)'=0.
\ena
Namely we find that the harmonic function can be, at most,
a sum of the functions
of $r$ and $u$. This is consistent with our previous result~\p{harm1} and gives
no further constraint.

Note that separable forms for the metric of the type~\p{sol} was assumed from
the beginning in \cite{CH,DMNT1,OP,DMNT2}, but here we have naturally derived this
property. Also the harmonic functions were taken to be independent of $u$,
but they can be actually functions of $u$ as well.

We still have to take Eq.~\p{fe2} into our account.
 This equation is rewritten as
\bea
W(u,r)+V(u)
+\half\sum_{\alpha=1}^{d-2}e^{2(\Xi-Z_\alpha)}\partial_\alpha^2 K
+\half e^{2(\Xi-B)}r^{-(\tilde{d}+1)}\left(
r^{(\tilde{d}+1)}K'
\right)'
=0
\label{last_eq}
\,,
\ena
where
\bea
&& W(u,r)\equiv
\sum_{A,B} \frac{(D-2)^2}{\Delta_A \Delta_B} (M_{AB}+2) (\ln H_A)^{\bm\cdot}
(\ln H_B)^{\bm\cdot}
+2 \sum_A \frac{D-2}{\Delta_A}(\ln H_A)^{\bm\cdot\bm\cdot}
\nn
&& \hs{20} +\; 4(D-2)(\dot \beta -\dot \xi)
\sum_A\frac{(\ln H_A)^{\bm\cdot}}{\Delta_A}
\,,
\label{def_W}
\\
&&
V(u) \equiv
\sum_{\a=1}^{d-2} \left(\ddot \zeta_{\a} + {\dot \zeta_{\a}}^2\right)
+ (\tilde d + 2)\left(\ddot \beta + {\dot \beta}^2\right)
- 2 \dot \xi \left[
\sum_{\a=1}^{d-2}
 \dot \zeta_{\a} + (\tilde d + 2) \dot \beta
\right]
+ \frac12 (\dot \phi)^2
\,.
~~~~~~
\label{def_V}
\ena

Eq.~(\ref{last_eq}) can be regarded formally as the equation for $K$,
which is an elliptic type differential equation with respect to
$r$ and $y^\alpha$. However the source terms depend not only on
$r$ but also on $u$. Hence we have to solve the
elliptic type  differential equation at any time $u$.
It may be very difficult to find the analytic solutions.
Instead we may first assume $K$
explicitly, and then solve Eq.~(\ref{last_eq}).
In this case, Eq.~(\ref{last_eq})  must be regarded as
a constraint equation for the formally solved variables
$\Xi,Z_\alpha, B$ and $\Phi$.
In this paper, we shall adopt the latter approach.

Here we assume
\bea
K = e^{-2\xi(u)}k(u,y^\a)+{m(u,y^\a)\over r^{\td}}
\label{K1}
\,,
\ena
with
\bea
k(u,y^\a)&=&
k_0(u) + \sum_{\a=1}^{d-2}
k_\a(u) y^\a
+\sum_{\a,\b=1(\a\neq \b)}^{d-2}k_{\a\b}(u) y^\a y^\b
+\sum_{\a \in ^\forall q_A} e^{2\zeta_\a(u)}
h_{\a\a}(u) (y^\a)^2,~~~~
\label{K1-1}
\\
m(u,y^\a)
&=&m_0(u) + \sum_{\a=1}^{d-2} m_\a(u) y^\a
\label{K1-2}
\,,
\ena
where $k_0(u),k_\a(u),k_{\a\b}(u), h_{\a\a}(u), m_0(u)$ and $m_\a(u)$
are arbitrary functions of $u$. Here the sum $\a \in ^\forall\!\! q_A$
is taken only over $y^\a$ coordinates belonging to all the branes.

Given $K(u,r,y^\a)$, we find that $u$-dependent terms
($\xi, \zeta_\a, \beta$ and $\phi$) are constrained by two
conditions~(\ref{cond1}) and (\ref{last_eq}).
The solution is then given by
\bea
ds_D^2 &=& \prod_A H_A^{2 \frac{q_A+1}{\Delta_A}}
\Bigg[ e^{2\xi(u)}
\prod_A H_A^{- 2 \frac{D-2}{\Delta_A}} \left(-2dudv +K(u,r,y^\a) du^2\right)
\nn
&& \hs{20} + \; \sum_{\a=1}^{d-2}
\prod_A H_A^{- 2 \frac{\c_A^{(\a)}}{\Delta_A}}
e^{2\zeta_{\a}(u)} (dy^\a)^2 +
e^{2\beta(u)}\left(dr^2 + r^2 d\Omega_{\tilde d+1}^2\right) \Bigg], \nn
&& \tilde E_A \color{black}= \sqrt{\frac{2(D-2)}{ \Delta_A}} H^{-1}_A, \quad
\Phi = \sum_{A} \epsilon_A a_A \frac{D-2}{\Delta_A} \ln H_A + \phi(u),
\label{res1}
\ena
with two constraints (\ref{cond1}) and (\ref{last_eq}), where
$H_A$ and $K$ are given by Eqs.~(\ref{harm1}) and (\ref{K1}), respectively, and
\bea
\c_A^{(\a)} = \left\{ \begin{array}{l}
D-2 \\
0
\end{array}
\right. \hs{5} {\rm for} \hs{3} \left\{
\begin{array}{l}
y^\a \mbox{   belonging  to $q_A$-brane} \\
{\rm otherwise}
\end{array}
\right.
\label{gamma}
\,.
\ena
Note that we still have one gauge freedom for the time coordinate $u$,
by which we can choose any function for $\xi(u)$.

To give solutions explicitly, Eqs.~(\ref{cond1}) and (\ref{last_eq})
must still be solved.
Let us now discuss explicit solutions.

\section{Solutions with time-independent harmonic functions}

To see the relation of our results with earlier work, let us first discuss
solutions with $u$-independent harmonic functions, i.e.,
\bea
H_A=h_A^{(0)}+\frac{Q_A}{r^{\td}},
\ena
where $h_A^{(0)}$ and $Q_A$ are constants.
In this case, since $\dot H_A=0$, we have $W=0$.

We now discuss two examples.

\subsection{Branes with factorized metrics in time and space}

If $h_{\a\a}(u) =0 (\a\in ^\forall q_A)$, the conditions on $u$-dependent terms
($\xi, \zeta_\a, \beta$ and $\phi$) should satisfy are Eq. (\ref{cond1}) and
Eq. (\ref{last_eq}) with $W=0$, i.e.
\bea
\sum_{\a=1}^{d-2} \left(\ddot \zeta_{\a} + {\dot \zeta_{\a}}^2\right)
+ (\tilde d + 2)\left(\ddot \beta + {\dot \beta}^2\right)
- 2 \dot \xi \left[
\sum_{\a=1}^{d-2}
 \dot \zeta_{\a} + (\tilde d + 2) \dot \beta
\right]
+ \frac12 (\dot \phi)^2=0
\,.
~~~~~~
\label{res2}
\ena

The solutions discussed in \cite{DMNT1,DMNT2} belong to this class.
They consider a single D3-brane with $d=4, \td =4$ and take
\bea
H_3 = \frac{R^4}{r^4}, ~~
e^{2\xi} = e^{2\zeta_{1}}  = e^{2\zeta_{2}}
\equiv e^{f(u)}, ~~
K = \beta=0, ~~
a_A=0,~~
\Phi=\phi(u)
\,,
\ena
where $R$ is a constant. The metrics here are of the factorized
form in $u$- and $r$-dependent terms.
Eq.~\p{cond1} is trivially satisfied, and Eq.~\p{res2} gives
\bea
\ddot f - \frac12 \dot f^2 + \frac12 \dot \phi^2 =0,
\ena
in agreement with their result. Here we have more general intersecting
solutions with the function $K$.

\subsection{Branes in pp-wave backgrounds}

Here, we give an example with the pp-wave,
\bea
K=e^{-2\xi(u)}k(u,y^\a)
\,,
\label{K3}
\ena
with (\ref{K1-1}). This is just the case with $m=0$ in Eq. (\ref{K1}).
The condition \p{last_eq} reduces to
\bea
V + \sum_{\a \in ^\forall q_A} h_{\a\a} = 0,
\label{res3}
\ena
where $V$ is defined by Eq.~\p{def_V}.
If this condition and Eq.~(\ref{cond1}) are satisfied,
the solution is Eq.~(\ref{res1}) with (\ref{K3}).

For a check, let us compare with the D3-brane solution in \cite{CH},
in which they have $d=4, \td =4$ and
\bea
&& H = \frac{R^4}{r^4}, ~~
e^{2\xi} \equiv k_{\rm CH}^2(u) ~,~~
e^{2\xi} K =  k(u,y^\a) \equiv h_{\rm CH}(u, r, y^\a)
, \nn
&&\zeta_{1}=\zeta_{2}=\beta=0, ~~
a_A=0,~~
\Phi=\phi(u)\equiv \phi_{\rm CH}(u),
\ena
where $k_{\rm CH}, h_{\rm CH}, \phi_{\rm CH}$ are
 the variables adopted in \cite{CH}.
Again Eq.~\p{cond1} is trivial and Eq.~\p{res3} reduces to
\bea
\frac12 \dot \phi^2 = - h_{11}- h_{22},
\ena
in agreement with Eq. (12) in \cite{CH}.

As a more interesting case, let us consider D1-D5-brane solution:
\bea
ds^2 &=& H^{-\frac{3}{4}}_1
H^{-\frac{1}{4}}_5 e^{2\xi(u)}\left[-2 du dv + K(u,y^\a) du^2\right] +
\left(\frac{H_1}{H_5}\right)^{\frac{1}{4}}
\sum^4_{\alpha =1} e^{\zeta_{\a}(u)} dy^2_\alpha \nn
&& + \; H^{\frac{1}{4}}_1 H^{\frac{3}{4}}_5
e^{2\beta(u)}\left(dr^2 + r^2 d\Omega^2_3\right), \nn \Phi
&=& \ln \left(\frac{H_1}{H_5}\right)^{\frac{1}{2}} + \phi(u).
\label{d1-d5}
\ena
In this case, $K$ depends on $y^\a$ linearly because
only one spatial dimension in $u$-$v$ coordinates can intersect,
and so we take
\bea
K = e^{-2\xi(u)}\left(k_0(u) + \sum_{\a=1}^{d-2} k_\a (u) y^\a
\right).
\ena
The conditions~\p{cond1} and \p{res3} tell us that
\bea
\phi = 4\beta = - \sum_{\a=1}^4 \zeta_{\a},~~
\sum_{\a=1}^{4} {\dot \zeta_{\a}}^2 + 12{\dot \beta}^2=0.
\ena
The last relation implies that $\zeta_{\a}$ and $\beta$ are constant,
giving no nontrivial solutions for these. The discussions in \cite{OP}
overlooked Eqs.~\p{fe3} and \p{fe8},
and so these restrictions on the solution were not obtained.

\section{Solutions with time-dependent harmonic functions}

In this section, we present more nontrivial solutions with both
$r$- and $u$-dependent harmonic functions $H_A$ in \p{harm1}.
These are the new solutions which have not been known.
The metric is also given by \p{res1} with \p{harm1} in the general case.
For simplicity of presentation, let us restrict ourselves to the simple
case of $K=0$. The nontrivial constraint we still have is the $uu$-component
of the Einstein equation~\p{last_eq}, which reduces to
\bea
W(u,r) + V(u) = 0.
\label{cond2}
\ena
As we discussed, we regard this equation as a constraint on
the time-dependent part of the harmonic and metric functions.
Note that we can easily extend the solution to non-vanishing $K$,
if $K$ is given by (\ref{K1}). In the case with the quadratic terms
of $y^\a$, the condition (\ref{cond2}) should be replaced by
\bea
W(u,r) + V(u) +\sum_{\a \in ^\forall q_A} h_{\a\a} = 0.
\label{cond3}
\ena
In what follows, we solve Eq.~(\ref{cond2}) and give
nontrivial solutions.

\subsection{Single brane}

We first consider a single $A$-brane.
For all branes in M-theory and superstrings, we have the relation
\bea
\frac{2(D-2)}{\Delta_A}=1,
\ena
so it is sufficient to concentrate on those solutions in which this relation
is valid. In this case, \p{cond2} gives
\bea
\frac{\ddot H_A}{H_A} + 2(\dot \beta -\dot \xi)\frac{\dot H_A}{H_A} +V=0.
\ena
Substituting \p{harm1} and sorting out the terms in the orders of $r$,
we find
\bea
\label{l1}
&&
\sum_{\a=1}^{d-2} \left(\ddot \zeta_{\a} + {\dot \zeta_{\a}}^2\right)
+ (\tilde d + 2)\left(\ddot \beta + {\dot \beta}^2\right)
- 2 \dot \xi \left[
\sum_{\a=1}^{d-2}
 \dot \zeta_{\a} + (\tilde d + 2) \dot \beta
\right]
+ \frac12 (\dot \phi)^2=0 ,
\\
\label{l2}
&& \ddot h_A +2(\dot \beta-\dot \xi) \dot h_A=0.
\ena
We can integrate \p{l2} to obtain
\bea
2(\beta-\xi) = -\ln \dot h_A -2 c_1,
\label{l21}
\ena
where $h_A$ is an arbitrary function of $u$, $c_1$ is an integration constant
and we have additional conditions~\p{cond1} as well as \p{l1}.

In the present solutions, we have $(d+2)$ arbitrary functions;
the metric functions
$\xi(u), \zeta_\a (u), \beta (u)$, the dilaton field $\phi(u)$,
and the gauge field $h_A(u)$.
We still have one constraint \p{cond1} and two equations \p{l1}
and \p{l21} for those variables.
Taking into account one gauge degree of freedom of $u$ coordinate,
there are $(d-2)$ degrees of freedom in the present single brane system.

As an example, let us consider D3-brane. In this case,
we have six unknown functions of $u$; $\xi, \zeta_1, \zeta_2, \beta$,
$\phi$, and $h_3$,
which must satisfy
\bea
\beta=0,~~
\xi = \frac12 \ln \dot h_3 + c_1,~~
\sum_{\a=1}^{2} \left(\ddot \zeta_{\a} + {\dot \zeta_{\a}}^2\right)
- \frac{\ddot h_3}{\dot h_3} \sum_{\a=1}^{2} \dot \zeta_{\a}
+ \frac12 \dot \phi^2 =0.
\ena
Using the gauge freedom, we may set $\xi=0$.
As a result, two functions (e.g.
$\zeta_1$ and $\zeta_2$) remain arbitrary.
This gauge choice reduces to $\dot h_3$=constant, that is
$h_3=au+b$.
If we assume $\zeta_1=\zeta_2\equiv f(u)/2$
and adopt the gauge condition such that $\xi=f(u)/2$, we find
the similar solution in \cite{DMNT1,DMNT2}, although $h_3$ depends on $u$.

\subsection{Intersecting two branes}
Let us consider two intersecting branes $A$ and $B$.
In this case, \p{cond2} gives
\bea
\frac{\dot H_A \dot H_B}{H_A H_B} +
\frac{\ddot H_A}{H_A}+\frac{\ddot H_B}{H_B}
+2\left(\dot \beta - \dot \xi\right) \left(\frac{\dot H_A}{H_A}
+\frac{\dot H_B}{H_B}\right)+V=0.
\ena
Substituting \p{harm1} and sorting out the terms in the orders of $r$,
we find
\bea
\label{m1}
&& V=0, \\
\label{m2}
&& Q_B \ddot h_A + Q_A \ddot h_B + 2\left(\dot \beta -\dot \xi\right)
 \left(Q_B \dot h_A + Q_A \dot h_B\right)
=0, \\
\label{m3}
&& \dot h_A \dot h_B + \ddot h_A h_B + h_A \ddot h_B
+2\left(\dot \beta -\dot \xi\right)
\left(\dot h_A h_B + h_A \dot h_B\right)=0.
\ena
To solve the last two equations, we introduce new variables
$f_\pm(u)$ as
\bea
h_A(u) &=& \frac{Q_A}{2} [f_+(u) +f_-(u)], \nn
h_B(u) &=& \frac{Q_B}{2} [f_+(u) -f_-(u)].
\label{sol_h12}
\ena
Eqs. (\ref{m2})and (\ref{m3}) are written as
\bea
\label{X1}
&& \ddot f_+ + 2\left(\dot \beta -\dot \xi\right)
 \dot f_+
=0, \\
\label{X2}
&& (\dot f_+)^2 =2f_-[\ddot f_-+2(\dot \beta -\dot \xi)
\dot f_-]+(\dot f_-)^2.
\ena
Integrating these equations, we find
\bea
\label{sol1}
&&  e^{2(\beta-\xi)}\dot f_+=c_+\,,
\\
\label{sol2}
&& \sqrt{f_-(f_- - c_-)}+c_-\ln \left[
\sqrt{f_-}+\sqrt{f_- - c_-}
\right]=f_++c_0
\,,
\ena
where $c_0$, $c_\pm$ are integration constants.

Once
we know $\beta(u)$, fixing the gauge
(i.e., giving $\xi(u)$), we can solve Eq. (\ref{sol1})
to obtain $f_+(u)$.
Then $f_-(u)$ is obtained by solving
Eq. (\ref{sol2}).
For example, if we choose the gauge as $\xi=\beta$,
we find
\bea
f_+(u)=c_+  u +d_+
\label{sol_fp}
\,,
\ena
where $d_+$ is an integration constant.
Then $f_-(u)$ is
implicitly given by the following equation;
\bea
\sqrt{f_-(f_- - c_-)}+c_-\ln \left[
\sqrt{f_-}+\sqrt{f_- - c_-}
\right]=
c_+  u +d_+
\,,
\label{sol_fm}
\ena
where we have chosen $c_0=0$ without loss of generality.

Compared with the single brane system, our intersecting brane system has
one additional function $h_B(u)$. On the other hand,
there is one additional constraint from \p{cond1} of the additional brane
as well as Eq.~\p{m1}.
As a result, naively we expect that $(d-3)$ degree of freedom
will be left in the present system. However, there are some exceptional cases.
If the number of intersecting dimensions is one,
e.g. for D1-D5, D2-D4, and D3-D3 intersecting brane systems,
the conditions~\p{cond1} and \p{m1} yield that all
arbitrary functions vanish, i.e.,
$\zeta_{\a}=\beta=\phi=0$.
We can set $\xi=0$ by use of the gauge freedom.
As a result no degree of freedom is left in those systems.

We show one concrete example, i.e., the D1-D5-brane system.
The solution is given by
\bea
ds^2 &=& -2 H^{-\frac{3}{4}}_1 H^{-\frac{1}{4}}_5  du dv +
\left(\frac{H_1}{H_5}\right)^{\frac{1}{4}}
\sum^4_{\alpha =1}dy^2_\alpha
+ H^{\frac{1}{4}}_1 H^{\frac{3}{4}}_5
\left(dr^2 + r^2 d\Omega^2_3\right), \nn
\Phi &=& \ln \left(\frac{H_1}{H_5}\right)^{\frac{1}{2}}.
\label{d1-d5 new}
\,
\ena
where $H_A$ is given by Eq. (\ref{harm1}) with $A=1$ or 5
and $\td=2$.
We have  chosen $\xi=0$ by
 using gauge degree of freedom.
Then $h_1(u)$ and $h_5(u)$ are given by Eq. (\ref{sol_h12}), i.e,
\bea
h_1(u) &=& \frac{Q_1}{2} [f_+(u) +f_-(u)], \nn
h_5(u) &=& \frac{Q_5}{2} [f_+(u) -f_-(u)]
\label{sol_h15}
\,,
\ena
where $f_+(u)$ is given by Eq. (\ref{sol_fp})
and
$f_-(u)$ is determined by inverting Eq. (\ref{sol_fm}) as a function of $u$.
Hence the solution is completely fixed up to some integration constants.

Next let us consider D2-D6-brane solution:
\bea
ds^2 &=&  H^{-\frac{5}{8}}_2 H^{-\frac{1}{8}}_6
\left( -2e^{2\xi(u)} du dv +e^{2\zeta_{1}(u)}(dy^1)^2\right)
+H^{\frac{3}{8}}_2 H^{-\frac{1}{8}}_6\sum^5_{\alpha =2}
e^{2\zeta_{\a}(u) }dy^2_\alpha \nn
&&+ H^{\frac{3}{8}}_2 H^{\frac{7}{8}}_6e^{2\beta(u)}
\left(dr^2 + r^2 d\Omega^2_2\right), \nn
\Phi &=& \frac{1}{4}\ln H_2 - \frac{3}{4}\ln H_6 + \phi(u)
\,.
\label{d2-d6}
\ena
In this case, there are ten non-trivial $u$-dependent functions
$\xi, \zeta_{\a} (\a=1\sim 5), \beta$,  $\phi$
and $f_\pm$ which must satisfy \p{cond1}, \p{res2}, \p{sol1}
and \p{sol2}:
\bea
&& \sum_{\a=2}^5\zeta_{\a}=-\phi =-\frac{4}{3}\beta\color{black},
\label{constraint1}
\\
&& \left( \zeta_{1} + \frac{5}{3}\beta \right)^{\hskip -.5em \ddot{~}} -
 2\dot \xi\left(\zeta_{1}+\frac{5}{3}\beta
\right)^{\hskip -.5em \dot{~}}
+\dot \zeta_{1}^2 +\frac{17}{3}\dot \beta^2
- \sum_{\a,\b=2}^5\dot \zeta_{\a}
\dot \zeta_{\b} (1-\d_{\a\b})=0,
~~~~~~~~~
\label{constraint2}
\\
&& \beta-\xi=-\half \left(\ln \dot f_+-\ln c_+
\right)\,,
\label{constraint3}
\\
&& \sqrt{f_-(f_- - c_-)}+c_-\ln \left[
\sqrt{f_-}+\sqrt{f_- - c_-}
\right]=f_++c_0
\,,
\label{constraint4}
\ena
where $c_0, c_\pm$ are arbitrary constants.
We can set $\xi=0$ by use of the gauge freedom.
As a result, we find four arbitrary functions.
For example,
one can take $\zeta_{2} , \zeta_{3} , \zeta_{4}$ and
$f_-$ to be arbitrary functions, and determine
$\zeta_{1}, \zeta_{5}, \beta$, $\phi$ and $f_+$
by using the five equations (\ref{constraint1}), (\ref{constraint2}),
(\ref{constraint3}) and (\ref{constraint4}).

We can easily extend these solutions to the cases with the
$K$-wave if $K$ is given in the form of (\ref{K1}).

\section{Some Properties of the Solutions}

We discuss some properties of our solutions.
There are three important geometrical properties of spacetime:
a singularity, a horizon, and an asymptotic structure.

\subsection{Singularity}

To study the spacetime singularity, we have to analyze the curvature tensors.
If matter fields are singular at some spacetime region, the Ricci curvature
will diverge. For the form and dilaton fields, we have
\bea
F_{n_A}^2&=& -n_A!\,e^{-\e_A a_A \Phi-2B}\left({H_A'\over H_A}\right)^2
\nn
&=&-n_A!\,e^{-\e_A a_A \phi-2\beta}\prod_B
H_B^{-\epsilon_A a_A \epsilon_B a_B
 \frac{D-2}{\Delta_B}}\prod_C H_C^{-2
{(q_C+1)\over \Delta_C}}\left({H_A'\over H_A}\right)^2,
\\
e^\Phi&=&e^{\phi(u)}\prod_A H_A^{\epsilon_A a_A \frac{D-2}{\Delta_A}}.
\label{dil2}
\ena
The Ricci scalar is given by
\bea
{\cal R}&=&\half (\pa \Phi)^2 +\half \sum_A\frac{2n_A-D}{D-2} e^{-2B}
\left({H_A'\over H_A}\right)^2
\label{scalarR}
\,.
\eea
Since the harmonic function $H_A$ diverges at $r=0$ if
the charge $Q_A$ does not vanish, we naively expect
that matter fields will diverge at $r=0$ as well.
However there are some exceptional cases in which
matter fields are regular even at $r=0$.
We can explicitly show it.

For a single brane system, we find that the second term of the
Ricci scalar behaves
as
\bea
{\rm the~second~term~of}~{\cal R}\propto r^{\frac{2(q_A+1)(7-q_A)}{\Delta_A}-2},
\ena
as $r\rightarrow 0$.
It diverges except for $q_A = 3$ (D3-brane).
In the case of the D3-brane, the dilaton coupling vanishes ($a_A=0$).
The dilaton $\Phi(=\phi(u))$ depends only on $u$ from Eq.~\p{dil2},
and then the first term of the Ricci scalar does not diverge either.

We can also calculate the Kretschmann invariant. We find
\bea
{\cal R}_{\mu\nu\rho\sigma}{\cal R}^{\mu\nu\rho\sigma}
\propto
r^{-\frac{(q_A-3)^2}{4}}
\,,
\eea
as $r\rightarrow 0$. Hence we again find that
D3-brane system is regular even at $r=0$, where branes exist.
However other single-brane system has a singularity at $r=0$.

For the D3-brane system, the Kretschmann invariant is given by
\be
{\cal R}_{\mu\nu\rho\sigma}{\cal R}^{\mu\nu\rho\sigma}
=\frac{80Q_3^2\left(Q_3^2+12r^8h_3^2\right)}{\left(Q_3+r^4h_3\right)^5} \, ,
\ee
where $h_3=a_3u+b_3$ with $a_3$ and $b_3$ being constants.
In that case, $r=0$ is not singular and
the curvatures are time independent.

However there appears  a singularity at
\be
r^4=r_s^4(u)\equiv -{Q_3\over a_3u+b_3} \, .
\ee
The position of this singularity is time-dependent unless $a_3=0$.
If $Q_3(a_3u+b_3)<0$, i.e. $u<-b_3/a_3$ (for $a_3 Q_3>0$)
or $u>-b_3/a_3$ (for $a_3 Q_3<0$) the singularity appears at $r_s>0$.
The Ricci scalar also diverges at the same spacetime position.
Hence even if $r=0$ is regular, there appears a singularity in the region of
$r>0$ either in the future or on the past.
Even if $r_s^4(u)<0$, because $r=0$ is not singular,
we may be able to extend the spacetime beyond $r=0$.
Then the singularity appears at $\tilde r^4\equiv
r^4+{Q_3/(a_3u+b_3)}=0$.
The regular brane at $r=0$ is static, but
the singularity is moving.

Note that if $a_3=0$, setting $b_3=1$ and introducing new radial coordinate
$\tilde r$ by $\tilde r^4=r^4+{Q_3}$, we find the metric as
\bea
ds_{10}^2&=&\left(1 -{Q_3\over \tilde r^4}\right)^{1/2}
\left[
-2dudv +K(u) du^2 +e^{2\zeta_1(u)}(dy^1)^2+
e^{2\zeta_2(u)}(dy^2)^2
\right]
\nn
&+&\left(1-{Q_3\over \tilde r^4}\right)^{-2}
d\tilde r^2+ \tilde r^2d\Omega_5^2
\,.
\ena
This is almost the same as the static D3-brane solution,
although there are three time-dependent arbitrary functions,
$K, \zeta_1$ and $\zeta_2$.
In this case, not only the regular brane at $\tilde r=\sqrt[4]{\mathstrut Q_3}$
is static\footnote[1]{Here we assume that $Q_3>0$. If $Q_3<0$, then
the singularity appears at $r=\sqrt[4]{-Q_3}(>0)$.}
but also the singularity at $\tilde r=0$ is time independent.

The analysis of an intersecting two-brane system is similar. The second term in
the Ricci scalar is proportional to
\be
r^{2\left[7+\bar q -(q_A+q_B)\right]\left[\frac{q_A+1}{\Delta_A}
+\frac{q_B+1}{\Delta_B}\right]-2}\,,
\ee
in the limit of $r=0$.
It  diverges except for the cases of $q_A+q_B=6$ and $\bar q =1$,
that is, D1-D5, D2-D4 and D3-D3 intersecting-brane systems.
In these cases, the dilaton couplings satisfy $a_A+a_B=0$,
hence the dilaton \p{res1} is constant near $r=0$.
Therefore the Ricci scalar does not diverge.
Calculating the Kretschmann invariant in
the intersecting two-brane system,
we find that $r=0$ is not singular in the D1-D5,
D2-D4 and D3-D3 intersecting brane cases, but it  is singular
for other intersecting two-brane systems.
For D1-D5, D2-D4 and D3-D3-brane systems,
we find
\bea
{\cal R}_{\mu\nu\rho\sigma}{\cal R}^{\mu\nu\rho\sigma}
&=&\frac{24Q_1^4Q_5^4+{\cal O}(r^2)}{(Q_1+r^2h_1)^{9/2}(Q_5+r^2h_5)^{11/2}}\,,
\\
{\cal R}_{\mu\nu\rho\sigma}{\cal R}^{\mu\nu\rho\sigma}&=&
\frac{24Q_2^4 Q_4^4+{\cal O}(r^2)}{(Q_2+r^2 h_2)^{19/4}(Q_4+r^2 h_4)^{21/4}}\,,
\\
{\cal R}_{\mu\nu\rho\sigma}{\cal R}^{\mu\nu\rho\sigma}&=&
\frac{24Q_3^4 Q_{\tilde{3}}^4+{\cal O}(r^2)}{(Q_3+r^2 h_3)^5
(Q_{\tilde{3}}+r^2
{h}_{\tilde{3}})^5}
\, ,
\eea
respectively.
Hence the spacetime structure at $r=0$ is regular,
and the spacetime near the branes is static.

The singularities appear at $r^2=r^2_{s+}(u)$
and $r^2=r^2_{s-}(u)$, where
$r^2_{s\pm}$ satisfy
\bea
&&r_{s\pm}^2(u) \equiv
-{2\over  f_+(u) \pm  f_-(u) }=-{2\over  c_+u+d_+\pm f_-(u) }
\,.
\eea
with $f_-(u)$ given by Eq. (\ref{sol_fm}).

\begin{figure}
\begin{center}
\includegraphics{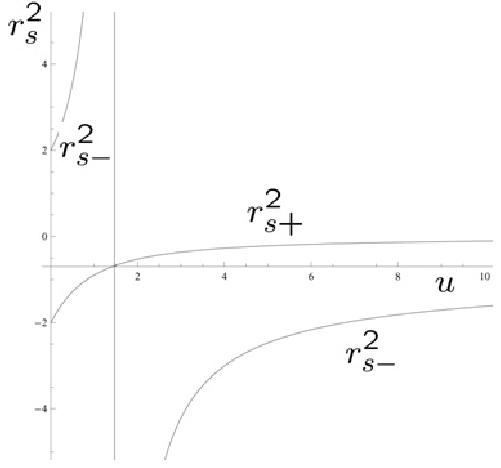}
\hskip 2cm
\includegraphics{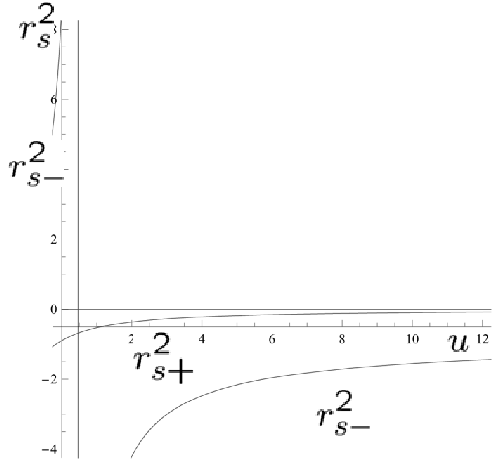}
\end{center}
\caption{\footnotesize{The position ($r_s^2$) of the singularity
with respect to time $u$ for D1-D5-brane system.
We set $c_+=1, d_+=0, c_-=1$ (left) and
$c_+=1, d_+=1, c_-=1$ (right).} }
\label{fg:fig1}
\end{figure}

The position of the singularity as $u$ changes is depicted in Fig.~\ref{fg:fig1}.
If $c_+=1, d_+=0, c_-=1$, $r^2_{s+}$ evolves into the $r^2>0$ region,
which means that the singularity appears beyond the regular position.
It can be a naked singularity. On the other hand,
if $c_+=1, d_+=1, c_-=1$, $r^2_{s+}$ does never go beyond the brane position
($r^2=0$). $r^2_{s-}$ is also behind $r=0$ for $u>u_0 (\approx 0.5)$.
Hence the singularity is covered by the regular branes.

\subsection{Spacetime structure near branes and horizons}

As we showed, the systems of single D3-brane,
D1-D5, D2-D4 and D3-D3 intersecting two branes
are regular at the position of branes ($r=4$).
The spacetime structure there is also static.

Assuming $K=0$ and choosing $\xi=0$, if we take the limit of $r=0$ in D3-brane,
we find
\bea
ds_{10}^2 &=& Q_3^{1/2}
\Bigg[{1\over z^2}
\left(
-2dudv +
dz^2
 \right) +  d\Omega_5^2
\Bigg]+ \; \sum_{\a=1}^{2}
e^{2\zeta_{\a}(u)} (dy^\a)^2
\label{near_D3}
\,,
\ena
where $z^2=Q_3/r^2$, and
$\xi$ is eliminated by the choice of gauge.
We find AdS$_3\times S^5\times \tilde E^2$,
where $\tilde E^2$ is a 2-dimensional
time-dependent flat Euclidean space.

In the case of D1-D5, D2-D4, D3-D3 intersecting branes,
we find
\bea
ds_{10}^2 &=&
Q_1^{1/4}Q_5^{3/4}\Bigg[{1\over z^2}\left( -2dudv +
dz^2\right) +  d\Omega_3^2 \Bigg]
+ \;\left({Q_1\over Q_5}\right)^{1/4} \sum_{\a=1}^{4} (dy^\a)^2,
\label{near_D1D5}
\nn
ds_{10}^2 &=&
Q_2^{3/8}Q_4^{5/8}\Bigg[{1\over z^2}\left( -2dudv +
dz^2\right) +  d\Omega_3^2 \Bigg]
+ \;\left({Q_4\over Q_2}\right)^{5/8} (dy^1)^2+
\left({Q_2\over Q_4}\right)^{3/8}\sum_{\a=2}^{4} (dy^\a)^2,
\label{near_D2D4}
\nn
ds_{10}^2 &=&
Q_3^{1/2}Q_{\tilde 3}^{1/2}\Bigg[{1\over z^2}\left( -2dudv +
dz^2\right) +  d\Omega_3^2 \Bigg]
+ \;\left({Q_{\tilde 3}\over Q_3}\right)^{1/2}\sum_{\a=1}^{2} (dy^\a)^2
+\left({Q_3\over Q_{\tilde 3}}\right)^{1/2}
\sum_{\a=3}^{4} (dy^\a)^2,
\label{near_D3D3}
\nn
\ena
where $z^2=Q_A Q_B/r^2$.
These spacetimes are AdS$_3\times S^3\times E^4$,
where $E^4$ is 4-dimensional flat Euclidean space.
This is a static spacetime.

As for the horizon, the event horizon can be easily
defined if the spacetime is  static.
However because our spacetime is time dependent,
it is not trivial. Rather
we may have to look for the
apparent horizon.
If $K=0$, our spacetime depends only on two variables $u$ and $r$,
then one may think that it is easy to find the apparent horizon
just as the analysis of
the apparent horizon in a spherically symmetric gravitational collapse.
However, because $u$ is a null coordinate which is not defined by $r$
but by another spatial coordinate in the brane worldvolume,
we have to analyze effectively
a three-dimensional problem to find the apparent horizon.
One may need numerical analysis, which is beyond the scope of the present study.

In the cases of a single D3-brane, or D1-D5, D2-D4,
and D3-D3 intersecting two-brane systems,
$r=0$ is not singular but regular and static.
Then $r=0$ could be an event horizon in 10 dimensional spacetime.
However, we cannot compactify one common brane direction
($y^{d-1}=(v-u)/\sqrt{\mathstrut 2}$),
and then we cannot obtain the lower-dimensional black holes

\subsection{Asymptotic structure}

As for the asymptotic spacetime structure, it is well known
if the spacetime is static.
In $(\td+3)$-dimensional spacetime, we find the asymptotically flat
Minkowski geometry, while in the brane directions ($(d-1)$-dimensional space),
we have a uniform and static geometry.
So compactifying all brane directions, we find an asymptotically flat
spacetime.

In the present time-dependent spacetime, we also find a uniform
geometry in the brane directions but it is time-dependent.
In $(\td+3)$-dimensional spacetime, in the limit of $
r\rightarrow +\infty$, we find
\bea
ds_D^2=-2dudv+\sum_{\a=1}^{d-2}f_\a(u)(dy^\a)^2+g(u)d\,{\vect r}_{\tilde d+2}^2
\,,
\ena
where we have used the gauge condition to set $g_{uv}=-1$, i.e.,
\bea
e^{2\xi}=\prod_Ah_A^{2(D-q_A-3)\over \Delta_A}
\,,
\ena
and $f_\a(u)$ and $g(u)$ are given by
\bea
f_\a(u) \equiv
 \prod_Ah_A^{-{2\delta_A^{(\alpha)}\over \Delta_A}}e^{2\zeta_\alpha}
\hspace{.5cm}{\rm and}
\hspace{.5cm}
g(u) \equiv
 \prod_Ah_A^{2(q_A+1)\over \Delta_A}e^{2\beta}
\,\,\,,
\ena
respectively.
If we compactify $y^\a$, we may find a time-dependent cosmological
solution. However, one spatial brane direction ($y^{d-1}$)
cannot be compactified, and then such a spacetime
is no longer a homogeneous FRW universe.
Rather it is just  a plane symmetric $(\td+3)$-dimensional inhomogeneous
spacetime,
unless we restrict ourselves to some position in the $y^{d-1}$ direction
just as a brane-world scenario.

\section{Supersymmetry}
\label{susy}

In this section, we explore the supersymmetry of our solutions.
The supersymmetry transformations in type II supergravities
 in the Einstein frame are
\bea
&& \delta \psi_\mu =
\Biggl[ \pa_{\mu} + \frac{1}{4} \omega_{\mu}^{\hat a \hat b }
\Gamma_{\hat a \hat b } + \frac{1}{8}\sum_{A} e^{\half \e_A a_A \Phi}
\left(1+\half \e_A a_A \right)
F\sr e^{\hat a}_{\mu}\Gamma_{\hat a}\,\,\Biggr]\e
\label{transgra} \,,\\
&& \delta \lambda = \left[ \partial\,\sr\Phi\,+\sum_{A}\frac{(-1)^{n_A}}{2} a_A
e^{\half\e_Aa_A\Phi}F\sr\,\,\right]\e
\label{transdila}
\,,
\ena
for the dilatino $\lambda$ and the gravitino $\psi_{\mu}$. The supersymmetry
parameter $\e$ is a Majorana (complex Weyl) spinor in type IIA (IIB) theory.
$\Gamma^{11}$
is given by
\bea
\Gamma^{11}\equiv \half \left(\Gamma^{\hat u}\Gamma^{\hat v}
-\Gamma^{\hat v}\Gamma^{\hat u}\right)\Gamma^{\hat y^1}\cdots
\Gamma^{\hat y^{d-2}}
\Gamma^{\hat r}\Gamma^{\hat \t_1}\cdots\Gamma^{\hat \t_{\td+1}}
\,,
\ena
and $\Gamma_{\hat a \hat b}
\equiv \Gamma_{[\hat a}\Gamma_{\smash{\hat b}]}$
are antisymmetrized gamma matrices.
The spin connection $\omega_{\mu \hat{a}\hat{b}}$ is defined by
\bea
\omega_{\mu \hat{a}\hat{b}}
\equiv \half e^{\nu}_{\hat{a}} ( \pa_{\mu}e_{\hat{b}\nu} - \pa_{\nu}e_{\hat{b}\mu} )
- \half e^{\nu}_{\hat{b}} ( \pa_{\mu}e_{\hat{a}\nu} - \pa_{\nu}e_{\hat{a}\mu} )
- \half e^{\rho}_{\hat{a}}e^{\sigma}_{\hat{b}}e^{\hat{c}}_{\mu}
(\pa_{\rho}e_{\hat{c}\sigma}-\pa_{\sigma}e_{\hat{c}\rho}),
\ena
where $e_{\mu}^{\hat a}$ is a vielbein satisfying
$e^{\hat a}_{\mu}e_{\hat a \nu}=g_{\mu\nu}$ and
$e^{\hat a}_{\mu}e^{\hat b \mu}=\eta^{\hat a \hat b}$.
Note that our Minkowski metric is given by
$\eta^{\hat u \hat v}=
\eta^{\hat v \hat u}=-1,\,
\eta^{\hat u \hat u}=\eta^{\hat v \hat v} =0$ and
$\eta^{\hat a \hat b } = \delta^{\hat a \hat b}$ for other indices,
because we use double null coordinates.
$F\sr$ denotes the R-R field contracted with gamma matrices;
e.g. $F\sr_3=\frac{1}{3!}F_{\mu\nu\rho}\Gamma^{\mu\nu\rho}$.
Similarly $\partial\,\sr\Phi\,= \Gamma^\mu \pa_\mu \Phi $.

We take $\e$ to be dependent on the coordinates $u$ and $r$ and
write $\e =s(u,r) \e_0$, where $\e_0$ is a constant spinor.
The Killing spinor equations are obtained by setting the above transformations
(\ref{transgra}) and (\ref{transdila}) to zero.
We find
\bea
\d \psi_u &=& \Biggl[ \frac{\dot{s}}{s} - \half\dot{\xi}
\left(1+\Gamma_{\hat{u}} \Gamma_{\hat{v}} \right)
- \half e^{\Xi-B}\,\Xi'\,\Gamma_{\hat{r}}\Gamma_{\hat{u}}\Biggr. \nn
&&\qquad +\frac{1}{4} e^{\Xi-B}(K'+K \Xi')\Gamma_{\hat r}\Gamma_{\hat v}+\frac{1}{4}\sum_{\a=1}^{d-2} e^{\Xi-Z_\a}\pa_\a K \Gamma_{\hat \a}\Gamma_{\hat v} \nn
&&\qquad \qquad +\; \frac{1}{8}e^{\Xi}\sum_{A} e^{\half \e_A a_A \Phi} \left(1+\half \e_A a_A \right)
F\sr \left(\Gamma_{\hat u}-\half K\Gamma_{\hat v}\right) \Biggr]\e_0=0,
\label{stu}\\
\d \psi_v &=& \Biggl[ - \half e^{\Xi-B}\,\Xi'\,\Gamma_{\hat{r}}\Gamma_{\hat{v}}
+\frac{1}{8}e^{\Xi}\sum_{A} e^{\half \e_A a_A \Phi} \left(1+\half \e_A a_A \right)
F\sr \Gamma_{\hat v} \Biggr]\e_0=0,
\label{stv}\\
\d \psi_{\a} &=& \Biggl[ \half e^{Z_{\a}-\Xi}\,
\dot{Z}_{\a}\,\Gamma_{\hat{v}}
\Gamma_{\hat{\a}} - \half e^{Z_{\a}-B}\,Z_{\a}'\,
\Gamma_{\hat{r}}\Gamma_{\hat{\a}} \nn
&&\qquad\qquad\qquad\qquad
+\; \frac{1}{8}e^{Z_{\a}}\sum_{A} e^{\half \e_A a_A \Phi}
\left(1+\half \e_A a_A \right)
F\sr \Gamma_{\hat \a} \Biggr]\e_0=0,
\label{sta}\\
\d \psi_r &=& \Biggl[ \frac{s'}{s} + \half e^{B-\Xi}\,\dot{B} \,\Gamma_{\hat{v}}
\Gamma_{\hat{r}}+\frac{1}{8}e^{B}\sum_{A} e^{\half \e_A a_A \Phi}
\left(1+\half \e_A a_A \right) F\sr \Gamma_{\hat r} \Biggr]\e_0=0,
\label{str}\\
\d \psi_{\t} &=& \Biggl[ \half e^{B-\Xi}\,\dot{B}\,r\,\,\Gamma_{\hat{v}}
\Gamma_{\smash{\hat{\t}}} - \half (1+r\,B')\,\,\Gamma_{\hat{r}}
\Gamma_{\smash{\hat{\t}}} \nn
&&\qquad\qquad\qquad\qquad
+\frac{1}{8}re^{B}\sum_{A} e^{\half \e_A a_A \Phi} \left(1+\half \e_A a_A \right)
F\sr \Gamma_{\smash{\hat \t}}  \Biggr]\e_0=0,
\label{sti} \\
\delta \lambda &=& \left[ \partial\,\sr\Phi\,+\sum_{A}\frac{(-1)^{n_A}}{2} a_A
e^{\half\e_Aa_A\Phi}F\sr\,\,\right]\e_0 =0.
\label{dilatino}
\ena
The transformations for other angular components are almost the same as the
$\t$ component and do not give any extra conditions.
Now we are going to examine the supersymmetry transformation for several
solutions.

\subsection{D3-brane system}

Let us first consider the time-dependent D3-brane solution in type IIB supergravity,
for which the dilaton coupling $a_A$ vanishes.
Using the self-duality condition,
we obtain
\bea
F\sr &=& \half\left[F\sr_5+*F\sr_5\right]
=\half\left[F_{uvy^1y^2r}\Gamma^{uvy^1y^2r}
+\tilde F_{\t_1\t_2\t_3\t_4\t_5}\Gamma^{\t_1\t_2\t_3\t_4\t_5}\right] \nn
&=& e^{-B}\frac{H_3'}{H_3}\left(\Gamma_{\hat u \hat v \hat y^1 \hat y^2 \hat r}
+\Gamma_{\hat \t_1 \hat \t_2 \hat \t_3 \hat \t_4 \hat \t_5}\right),
\ena
where we have used Eqs.~\p{cond1} and \p{res1}
to go to the second line.
It is easy to see that for the background~\p{res1}, the dilatino
variation gives the condition
\bea
\Gamma_{\hat v}\e_0=0.
\ena
The gravitino variation \p{stu} takes the form
\bea
\d \psi_u = \Bigg[ \frac{\dot{s}}{s}-\half\dot{\xi}
- \frac{1}{16}e^{\Xi-B}\frac{H_3'}{H_3}\Gamma_{\hat{u}\hat{r}}
\left(\Bigl(1-\Gamma_{\hat{u}\hat{v}\hat{y}^1\hat{y}^2}\Bigr)
+ \left(1-\Gamma_{\hat{\t}_1\hat{\t}_2\hat{\t}_3\hat{\t}_4\hat{\t}_5\hat{r}}
\right)\right)\Bigg]\e_0=0.
\label{d3gr}
\ena
The other conditions from \p{stv} -- \p{dilatino} are similar.
One can check that all these conditions are satisfied if $\e$ is given by
\bea
\e=H_3^{-\frac{1}{8}}(u,r) e^{\half \xi(u)}\e_0 \,,
\label{e_e0}
\ena
with
\be
\Gamma_{\hat{u}\hat{v}\hat{y}^1\hat{y}^2}\e_0=\e_0,\quad
{\rm and}~~~~
\Gamma_{\hat v}\e_0=0
\,.
\label{susy3}
\ee
Assuming the above conditions are satisfied, we find
the last term in Eq.~\p{d3gr} automatically vanishes.
The condition in~\p{e_e0} is needed to kill the $r$- and
$u$-dependent terms.
The first condition in \p{susy3} is the standard one for supersymmetry
in the presence of a D3-brane and leaves 16 supersymmetries unbroken.
We see that the second condition in \p{susy3} breaks the supersymmetry further
by half leaving 8 supersymmetries in total.
Namely, compared with the static brane solutions with 16 supersymmetries, it is broken
by further one half due to the additional $u$-dependence of the system.
All backgrounds of a single brane thus preserve 8 supersymmetries, though
we will find that D1-brane is an exceptional case, preserving
16 supersymmetries.

\subsection{Intersecting D1-D5-brane system}

Next we consider intersecting D1-D5-brane system in type IIB supergravity.
\bea
F\sr\,\, &=& {F\sr}_3 + e^{-\Phi}\color{black}*{F\sr}_7
= F_{uvr}\Gamma^{uvr} + \tilde F_{\t_1\t_2\t_3}\Gamma^{\t_1\t_2\t_3} \nn
&=& e^{-B-\half\Phi}\frac{H_1'}{H_1}\Gamma_{\hat u \hat v \hat r}
+e^{-B+\half\Phi}\frac{H_5'}{H_5}\Gamma_{\hat{\t}_1 \hat{\t}_2 \hat{\t}_3}.
\ena
The supersymmetry transformation of dilatino is
\bea
\d \lambda &=& \Biggl[ -\dot \Phi e^{-\Xi} \Gamma_{\hat v}
+ \Phi' e^{-B}\Gamma_{\hat r} - \half e^{\half\Phi} {F\sr}_3\color{black}
- \half e^{-\frac{3}{2}\Phi}*\! {F\sr}_7\color{black}\Biggr]\e_0 \nn
&=& \Biggl[ -\dot \Phi e^{-\Xi}\Gamma_{\hat v} +\half e^{-B} \frac{H_1'}{H_1}
\Gamma_{\hat r}\left(1-\Gamma_{\hat u \hat v}\right)-\half
e^{-B}\frac{H_5'}{H_5}
\Gamma_{\hat r}\left(1-\Gamma_{\hat{\t}_1\hat{\t}_2\hat{\t}_3\hat r}\right)
\Biggr]\e_0=0.
\label{di}
\ena
Using $\lambda\; (\Gamma^{11}\lambda=\lambda)$ and $\Gamma^{11}\e=-\e$, we find
\be
\Gamma_{\hat v}\e_0=0 , \quad
\left(1-\Gamma_{\hat u \hat v \hat y^1\hat y^2 \hat y^3 \hat y^4 }\right)
\e_0=0.
\ee
We now check the other condition from the gravitino:
\begin{align}
\d \psi_u &= \Biggl[\left(\frac{\dot s}{s} -\half\dot \Xi\right)
- \frac{1}{16}e^{\Xi-B}\Gamma_{\hat u \hat r}\left( 3\frac{H_1'}{H_1}
\left(1-\Gamma_{\hat u \hat v}\right) +\frac{H_5'}{H_5}
\left(1-\Gamma_{\hat{\t}_1\hat{\t}_2\hat{\t}_3\hat r}\right)\right)\Biggr]
\e_0=0,
\label{gr1} \\
\d \psi_{r}\color{black} &= \Biggl[\frac{s'}{s} + \frac{3}{16}\frac{H_1'}{H_1}
\Gamma_{\hat u \hat v} +\frac{1}{16}\frac{H_5'}{H_5}
\Gamma_{\hat{\t}_1\hat{\t}_2\hat{\t}_3\hat r}\Biggr]\e_0=0.
\end{align}
The other conditions are again similar. All these conditions are satisfied
if and only if
\be
\e =H_1^{-\frac{3}{16}}(u,r) H_5^{-\frac{1}{16}}(u,r)\,e^{\half \xi(u)}
\,\e_0 \, ,
\label{e_e01}
\ee
with
\be
\Gamma_{\hat u \hat v \hat y^1\hat y^2 \hat y^3 \hat y^4 }\e_0=\e_0,\quad
\Gamma_{\hat v}\e_0=0 .
\label{susy15}
\ee
At first glance, it seems that we have two conditions
$\left(1-\Gamma_{\hat u\hat v}\right)\e=0$
and $\Gamma_{\hat v}\e=0$ from \p{di} and \p{gr1}.
However it turns out from the property of gamma matrices that these two
conditions are equivalent.
Thus intersecting D1-D5-brane system also preserves eight supersymmetries.
This is a little surprising result because the number of the remaining
supersymmetry is the same as the single branes, but this is special for
the solutions involving D1-branes.

We also see from this analysis that we get twice the supersymmetries for
D1-brane compared with other single branes. The reason is that
two conditions coming from the  D1-brane and from the time dependence
degenerate to one ($\Gamma_{\hat v}\e_0=0$),
but those conditions are independent for the cases of other branes
as we have seen in the previous subsection.

\subsection{Intersecting D2-D6-brane system}

In this final subsection, let us consider the intersecting D2-D6-branes in
type IIA supergravity, although it is singular at the branes.

Now we have
\bea
F\sr &=& {F\sr}_4 + e^{-\frac{3}{2} \Phi}\color{black}*{F\sr}_8 =F_{uvy^1r}\Gamma^{uvy^1r}
+ \tilde F_{\t_1\t_2}\Gamma^{\t_1\t_2} \nn
&=& e^{-B-\frac{1}{4}\Phi}\frac{H_2'}{H_2}\Gamma_{\hat u \hat v \hat y^1\hat r}
+e^{-B+\frac{3}{4}\Phi}\frac{H_6'}{H_6}\Gamma_{\hat{\t}_1 \hat{\t}_2}.
\ena
The supersymmetry transformation of the dilatino is
\bea
\d \lambda &=& \Biggl[ -\dot \Phi e^{-\Xi} \Gamma_{\hat v}
+ \Phi' e^{-B}\Gamma_{\hat r} + \frac{1}{4} e^{\frac{1}{4}\Phi} {F\sr}_4\color{black}
+\frac{3}{4} e^{-\frac{9}{4}\Phi}*\! F\sr_8\color{black}\Biggr]\e_0 \nn
&=& \Biggl[ -\dot \Phi e^{-\Xi}\Gamma_{\hat v}
+ \frac{1}{4} e^{-B} \frac{H_2'}{H_2}\Gamma_{\hat r}
\left(1-\Gamma_{\hat u \hat v\hat y^1}\right)-\frac{3}{4} e^{-B}\frac{H_6'}{H_6}
\Gamma_{\hat r}\left(1-\Gamma_{\hat{\t}_1\hat{\t}_2\hat r}\right)\Biggr]\e_0=0.
\label{d2d6}
\ena
So $\e_0$ needs to satisfy the conditions
\be
\Gamma_{\hat v}\e_0=0 ,\qquad
(1-\Gamma_{\hat u \hat v \hat y^1})\e_0=0 , \qquad
(1-\Gamma_{\hat u \hat v \hat y^1 \hat y^2 \hat y^3 \hat y^4 \hat y^5})\e_0=0
\,.
\ee
The last term in \p{d2d6} automatically vanishes for given these conditions.
The other nontrivial conditions are
\bea
\d \psi_u &=& \Biggl[\left(\frac{\dot s}{s} -\half\dot \Xi\right)
- \frac{1}{32}e^{\Xi-B}\Gamma_{\hat u \hat r}\left(5\frac{H_2'}{H_2}
\left(1-\Gamma_{\hat u \hat v\hat y^1}\right) +\frac{H_6'}{H_6}
\left(1-\Gamma_{\hat{\t}_1\hat{\t}_2\hat r}\right)\right)\Biggr]\e_0=0, \\
\d \psi_r &=& \Biggl[\frac{s'}{s} + \frac{5}{32}\frac{H_2'}{H_2}
\Gamma_{\hat u \hat v\hat y^1} +\frac{1}{32}\frac{H_6'}{H_6}
\Gamma_{\hat{\t}_1\hat{\t}_2\hat r}\Biggr]\e_0=0.
\ena
The other components are similar. We find that all these conditions are
 satisfied
if and only if
\be
\e= H_2^{-\frac{5}{32}}(u,r)H_6^{-\frac{1}{32}}(u,r)
e^{\half \xi(u)}\e_0
\label{e_e02}\,,
\ee
with
\be
\Gamma_{\hat u \hat v \hat y^1}\e_0=\e_0 ,\quad
\Gamma_{\hat u \hat v \hat y^1 \hat y^2 \hat y^3 \hat y^4 \hat y^5}\e_0=\e_0 ,\quad
\Gamma_{\hat v}\e_0=0
\,.
\label{susy26}
\ee
The first condition in \p{susy26} comes from the D2-brane,
the second one from the D6-brane, the last one from the dilaton.
The $r$- and $u$-dependence in (\ref{e_e02}) is needed to kill
the $r$- and $u$-dependent terms.
Thus intersecting D2-D6-brane system preserves 4 supersymmetries.

\section{Concluding Remarks}

In this paper we have constructed a fairly general family of
time-dependent intersecting brane solutions. An important property
of these solutions is that they preserve partial supersymmetry.
This is important because this property assures that there will be
the corresponding dual field theories according to
the gauge/gravity correspondence.
The dual theories can be used, for example, to study nonperturbative region
of the gravity sector such as the behaviors of the theory close to
the singularity.

Our solutions include known D-brane solutions as well as M-branes in
time-dependent backgrounds, but the known ones were restricted to those
with factorized form of the metrics in time- and space-dependent functions.
We have certainly given such solutions but even these are more general.
We have also given more general spacetime-dependent solutions with
time-dependent harmonic functions where the dependences on
time and space cannot be factorized, but they are sum of such terms.
These are new class of solutions, and we have studied their singularities,
spacetime structure near branes and asymptotic structures.

It is possible that we may get more general solutions if we relax the condition
that $U$ depends only on $u$, but it is known that such a generalization
gives only non-BPS solutions already for static solutions~\cite{MO}.
So our solutions are expected to be most general BPS solutions with
spacetime-dependence.

We have also examined how many supersymmetries remain unbroken in our solutions.
The number of those are 8, 8 and 4 for a single brane other than D1,
D1-D5-brane and D2-D6-branes, respectively.
The 8 remaining supersymmetries on the single brane corresponds to
$N=4$ supersymmetry in 2 dimensions and $N=2$ in 4 dimensions.
To have remaining supersymmetry, the time dependence can come in
only through $u$. Due to this restriction, however, we cannot compactify
one worldvolume direction, preventing us from obtaining lower-dimensional
black holes, as discussed in sect.~5. It would be thus interesting to study
brane solutions with different time dependence, although we would loose
unbroken supersymmetry.

The near brane geometry is $AdS_3 \times S^5 \times \tilde E^2$ in
the single D3-brane system
($\tilde E^2$ : two-dimensional time-dependent flat Euclidean space),
or
$AdS_3 \times S^3 \times E^4$
($E^4$  : four-dimensional flat Euclidean space)
in the D1-D5, D2-D4, and D3-D3-brane systems.
As argued in \cite{MS}, the corresponding dual field theory would be
two-dimensional conformal field theory, now in time-dependent backgrounds.
It would be interesting to examine the dual field theory~\cite{dual}
and try to understand how the spacetime singularity is described in such a theory.
We hope that our construction of these general time-dependent solutions is
useful for further study of the singularities and other aspects of
the gravitational systems through this kind of dual theories.

Although our solutions include all kinds of brane solutions with RR,
NS-NS and eleven-dimensional four-form backgrounds, we have given explicit
and detailed discussions of solutions and their properties only for one or
two RR fields (D-branes).
It is interesting to study our solutions in more detail for the cases
including NS-NS field and more branes. It is important because in a static system,
we need more than two branes to obtain a black hole solution after compactification
to lower dimensions.

Finally though we have constructed the supersymmetric solutions, if we
are interested in the direct application (rather than studying singularities
and so on) of the solutions to our world, it may be also interesting to
consider solutions with dynamical supersymmetry breaking.

\section*{Acknowledgements}

 The work was supported in part by the Grant-in-Aid for
Scientific Research Fund of the JSPS No. 19540308, No. 20540283
and No. 06042, and by the Japan-U.K. Research Cooperative Program.



\begin{thebibliography}{10}

\bibitem{time1}
H.~Liu, G.~W.~Moore and N.~Seiberg,
  JHEP {\bf 0206} (2002) 045
  [arXiv:hep-th/0204168];
  JHEP {\bf 0210} (2002) 031
  [arXiv:hep-th/0206182].
\bibitem{time2}
A.~Hashimoto and S.~Sethi,
  Phys.\ Rev.\ Lett.\  {\bf 89} (2002) 261601
  [arXiv:hep-th/0208126].
\bibitem{S}
J.~Simon,
  JHEP {\bf 0210} (2002) 036
  [arXiv:hep-th/0208165].
\bibitem{CLO}
R.~G.~Cai, J.~X.~Lu and N.~Ohta,
  Phys.\ Lett.\ B {\bf 551} (2003) 178
  [arXiv:hep-th/0210206].
\bibitem{null}
N.~Ohta,
  Phys.\ Lett.\ B {\bf 559} (2003) 270
  [arXiv:hep-th/0302140].
\bibitem{bb1}
B.~Craps, S.~Sethi and E.~P.~Verlinde,
  JHEP {\bf 0510} (2005) 005
  [arXiv:hep-th/0506180].
\bibitem{bb2}
M.~Li,
  Phys.\ Lett.\ B {\bf 626} (2005) 202
  [arXiv:hep-th/0506260].
\bibitem{bb3}
M.~Li and W.~Song,
  JHEP {\bf 0510} (2005) 073
  [arXiv:hep-th/0507185].
\bibitem{bb4}
Y.~Hikida, R.~R.~Nayak and K.~L.~Panigrahi,
  JHEP {\bf 0509} (2005) 023
  [arXiv:hep-th/0508003].
\bibitem{bb5}
B.~Chen,
  Phys.\ Lett.\ B {\bf 632} (2006) 393
  [arXiv:hep-th/0508191].
\bibitem{bb6}
J.~H.~She,
  JHEP {\bf 0601} (2006) 002
  [arXiv:hep-th/0509067].
\bibitem{bb7}
B.~Chen, Y.~l.~He and P.~Zhang,
  Nucl.\ Phys.\ B {\bf 741} (2006) 269
  [arXiv:hep-th/0509113].
\bibitem{IKO}
T.~Ishino, H.~Kodama and N.~Ohta,
  Phys.\ Lett.\ B {\bf 631} (2005) 68
  [arXiv:hep-th/0509173].
\bibitem{bb8}
D.~Robbins and S.~Sethi,
  JHEP {\bf 0602} (2006) 052
  [arXiv:hep-th/0509204].
\bibitem{She:2005qq}
J.~H.~She,
  Phys.\ Rev.\ D {\bf 74} (2006) 046005
  [arXiv:hep-th/0512299].
\bibitem{bb10}
B.~Craps, A.~Rajaraman and S.~Sethi,
  Phys.\ Rev.\ D {\bf 73} (2006) 106005  [arXiv:hep-th/0601062].
\bibitem{CH}
C.~S.~Chu and P.~M.~Ho,
 JHEP {\bf 0604} (2006) 013 [arXiv:hep-th/0602054].
\bibitem{bb12}
S.~R.~Das and J.~Michelson,
 Phys.\ Rev.\ D {\bf 73} (2006) 126006 [arXiv:hep-th/0602099].
\bibitem{DMNT1}
S.~R.~Das, J.~Michelson, K.~Narayan and S.~P.~Trivedi,
 Phys.\ Rev.\ D {\bf 74} (2006) 026002 [arXiv:hep-th/0602107].
\bibitem{lin}
F.~L.~Lin and W.~Y.~Wen,
 JHEP {\bf 0605} (2006) 013  [arXiv:hep-th/0602124].
\bibitem{bb13}
E.~J.~Martinec, D.~Robbins and S.~Sethi,
 JHEP {\bf 0608} (2006) 025 [arXiv:hep-th/0603104].
\bibitem{bb14}
H.~Z.~Chen and B.~Chen,
 Phys.\ Lett.\ B {\bf 638} (2006) 74 [arXiv:hep-th/0603147].
\bibitem{bb15}
T.~Ishino and N.~Ohta,
 Phys.\ Lett.\ B {\bf 638} (2006) 105 [arXiv:hep-th/0603215].
\bibitem{NP}
R.~R.~Nayak and K.~L.~Panigrahi,
 Phys.\ Lett.\ B {\bf 638} (2006) 362 [arXiv:hep-th/0604172].
\bibitem{KO}
H.~Kodama and N.~Ohta,
 Prog. Theor. Phys. {\bf 116} (2006) 295 [arXiv:hep-th/0605179].
\bibitem{NPS}
R.~R.~Nayak, K.~L.~Panigrahi and S.~Siwach,
 Phys.\ Lett.\ B {\bf 640} (2006) 214
 [arXiv:hep-th/0605278].
\bibitem{OP}
N.~Ohta and K.~L.~Panigrahi,
  Phys.\ Rev.\  D {\bf 74} (2006) 126003
  [arXiv:hep-th/0610015].
\bibitem{KU}
H.~Kodama and K.~Uzawa,
  JHEP {\bf 0603} (2006) 053
  [arXiv:hep-th/0512104];\\
P.~Binetruy, M.~Sasaki and K.~Uzawa,
  arXiv:0712.3615 [hep-th];\\
K.~Maeda, N.~Ohta and K.~Uzawa,
  arXiv:0903.5483 [hep-th].
\bibitem{DMNT2}
S.~R.~Das, J.~Michelson, K.~Narayan and S.~P.~Trivedi,
  Phys.\ Rev.\  D {\bf 75} (2007) 026002
  [arXiv:hep-th/0610053].
\bibitem{BPRW}
J.~Bedford, C.~Papageorgakis, D.~Rodriguez-Gomez and J.~Ward,
  arXiv:hep-th/0702093.
\bibitem{BC}
B.~Craps,
  Class.\ Quant.\ Grav.\  {\bf 23} (2006) S849
  [arXiv:hep-th/0605199].
\bibitem{OPS}
N.~Ohta, K.~L.~Panigrahi and S.~Siwach,
 Nucl.\ Phys.\ B {\bf 674} (2003) 306
 [Erratum-ibid.\ B {\bf 748} (2006) 309] [arXiv:hep-th/0306186].
\bibitem{DSB}
R.~Argurio, M.~Bertolini, S.~Franco, and S.~Kachru,
 JHEP {\bf 0701} (2007) 083 [arXiv:hep-th/0610212];
  JHEP {\bf 0706} (2007) 017 [arXiv:hep-th/0703236].
\bibitem{NO}
 N.~Ohta,
 Phys.\ Lett.\ B {\bf 403} (1997) 218 [arXiv:hep-th/9702164].
\bibitem{MO}
  Y.~G.~Miao and N.~Ohta,
  Phys.\ Lett.\  B {\bf 594} (2004) 218
  [arXiv:hep-th/0404082].
\bibitem{AEH}
  R.~Argurio, F.~Englert and L.~Houart,
  Phys.\ Lett.\  B {\bf 398} (1997) 61
  [arXiv:hep-th/9701042].
\bibitem{PT}
G.~Papadopoulos and P.~K.~Townsend,
  Phys.\ Lett.\  B {\bf 380} (1996) 273
  [arXiv:hep-th/9603087].
\bibitem{T}
A.~A.~Tseytlin,
  Nucl.\ Phys.\  B {\bf 475} (1996) 149
  [arXiv:hep-th/9604035].
\bibitem{MS}
 J.~M.~Maldacena and A.~Strominger,
 JHEP {\bf 9812} (1998) 005 [arXiv:hep-th/9804085].
\bibitem{dual}
C.~S.~Chu and P.~M.~Ho,
  JHEP {\bf 0802} (2008) 058
  [arXiv:0710.2640 [hep-th]];\\
M.~R.~Gaberdiel and I.~Kirsch,
  JHEP {\bf 0704} (2007) 050
  [arXiv:hep-th/0703001]
\end{thebibliography}
\end{document}